\documentclass{aa}
\input{psfig}

\catcode`\@=11 
\def\gsim{\ifmmode{\mathrel{\mathpalette\@versim>}}
\else{$\mathrel{\mathpalette\@versim>}$}\fi}
\def\lsim{\ifmmode{\mathrel{\mathpalette\@versim<}}
\else{$\mathrel{\mathpalette\@versim<}$}\fi} 
\def\@versim#1#2{\lower2.9truept \vbox{\baselineskip 0pt \lineskip 0.5truept
\ialign{$\m@th#1\hfil##\hfil$\crcr#2\crcr\sim\crcr}}} 
\catcode`\@=12

\def\mincir{\raise -2.truept\hbox{\rlap{\hbox{$\sim$}}\raise5.truept
\hbox{$<$}\ }}
\def  \magcir{\raise -2.truept\hbox{\rlap{\hbox{$\sim$}}\raise5.truept
\hbox{$>$}\ }} 

\begin{document}
\thesaurus{11.01.2, 11.09.1 M81, 11.19.1, 13.25.2}

\title{The low luminosity AGN in the LINER galaxy M81:
$BeppoSAX$ discovery of highly ionized gas}
\author{S. Pellegrini$^1$, M. Cappi$^2$, L. Bassani$^2$, G. Malaguti$^2$, 
G. G. C. Palumbo$^1$, M. Persic$^3$}
\offprints{S. Pellegrini, pellegrini@bo.astro.it}
\institute{
$^1$ Dipartimento di Astronomia, Universit\`a di Bologna,
via Ranzani 1, I-40127 Bologna; pellegrini@bo.astro.it, 
ggcpalumbo@bo.astro.it\\
$^2$ Istituto TeSRE/CNR, via Gobetti 101, I-40129 Bologna;
mcappi@tesre.bo.cnr.it, bassani@tesre.bo.cnr.it, malaguti@tesre.bo.cnr.it \\
$^3$ Osservatorio Astronomico di Trieste, via G.B. Tiepolo 11, I-34131 Trieste;
persic@sissa.it}
\date{Received 21 September 1999 / Accepted 1 November 1999}
\authorrunning{S. Pellegrini et al.}
\titlerunning{The low luminosity AGN in the LINER galaxy M81}
\maketitle
\begin{abstract}
The LINER nucleus of the nearby spiral galaxy M81 was pointed by {\it
BeppoSAX}, which caught it at the highest (2--10) keV flux level
observed so far.  The LECS, MECS and PDS data, extending over
(0.1--100) keV, are used to investigate the physical similarities and
differences between LINERs and AGNs.  The continuum is well fitted by
a power law of photon index $\Gamma \sim 1.84$, modified by little
absorption due to cold material; this extends to (0.1--100) keV the
validity of a similar $ASCA$ result.  Superimposed on the continuum
$BeppoSAX$ detects a 6.7 keV emission line (confirming another $ASCA$
result) and an absorption edge at $\sim 8.6 $ keV. Both spectral
features are consistent with being produced by iron at the same high
ionization level, and probably also with the same column density. So,
we suggest that they originate from transmission through highly
ionized thin material.
Concerning the origin of the continuum emission, we do not
observe signs of reflection from the optically thick material of an
accretion disk, as usually found in Seyfert 1's (a 6.4 keV emission
line and a broad bump peaking at 10--20 keV). The low bolometric
luminosity of the nucleus of M81 is consistent with being produced by
advection dominated accretion; in this case the X-ray emission should
be dominated by Comptonization, rather than by bremsstrahlung, in
order to reproduce the steep spectrum observed over the (0.1--100) keV
band.

\end{abstract}

\section{Introduction}

M81 (NGC3031) is a nearby spiral galaxy with a prominent
bulge and well defined spiral arms, morphologically similar to M31
(Table 1).  It has been well studied at all frequencies, from radio
(e.g., Beck et al. 1985, Bietenholz et al. 1996), to optical and UV
(e.g., Ho et al. 1996), to X-rays (Sect. 2).  On the basis of these
observations, M81 turned out to be the closest galaxy to show the
spectroscopic signatures of a LINER (Ho et al. 1997).  In addition,
some observed properties make the nucleus of M81 a good candidate for
a low luminosity AGN (LLAGN). These include the presence of a broad
component of the $H\alpha$ emission line (Peimbert \& Torres-Peimbert
1981), a compact radio core (Bietenholz et al. 1996), a pointlike
X-ray source coincident with the optical nucleus (Fabbiano 1988,
hereafter F88), and the presence of a power law continuum in the 2--10
keV energy band (Ishisaki et al.  1996, hereafter
I96).  Dynamical studies also suggest the presence of a super-massive
object at the galaxy nucleus, of mass $4\times 10^6 M_{\odot}$ (Ho
1999).

Here we report the results of an analysis of the properties of the
nucleus of M81 over 0.1--100 keV, and discuss them in relation to
previous X-ray observations (mostly in the 2--10 keV band) and the
unsolved issue of sorting out the physical similarities and
differences between LINERs and AGNs. In particular the origin of
LINERs is still uncertain: their emission line spectrum could be
powered by a LLAGN, in which case they could represent the missing
link between normal galaxies and the less luminous AGNs, or by
starburst activity (Terlevich et al. 1992).  The very large energy
band of {\it BeppoSAX} coupled to its spectral resolution (8\%
FWHM at 6 keV) is especially suited for reliably measuring the high
energy continuum, and to search for a thermal component and Fe-K
features.  Therefore we can investigate more accurately than with
previous instruments the similarities and differencies of the X-ray
properties of LINERs with those of AGNs.

\begin{table*}
\caption[] { General characteristics of M81}
\begin{flushleft}
\begin{tabular}{ l  rl  l  l l  l  l l  l l  l }
\noalign{\smallskip}
\hline
\noalign{\smallskip}
Type$^a$ & RA  & Dec & d$\, ^b$ &$B_{\rm T}^0 \, ^a$& log$L_{\rm B} \, ^c$ &
  $N_{\rm H,Gal} \, ^d$ \\
        &(J2000) &(J2000) &  (Mpc)     &   (mag)           & ($L_{\odot}$) &
  (cm$^{-2}$)        \\
\noalign{\smallskip}
\hline
\noalign{\smallskip}
 Sab & $9^h 55^m 33^s \hskip -0.1truecm .2$ & $69^{\circ} 03^{\prime} 55^{\prime\prime} $ & 3.6 &
 7.39 & 10.32  & 4.1$\times 10^{20}$ \\
\noalign{\smallskip}
\hline
\end{tabular} 
\end{flushleft}
\bigskip

$^a$ from de Vaucouleurs et al. (1991).  $B_{\rm T}^0$ is the total B magnitude,
 corrected for Galactic and internal extinction.

$^b$ Cepheid-based distance (Freedman et al. 1994).

$^c$ total B-band luminosity, derived using the indicated distance and 
$B_{\rm T}^0$. 

$^d$ Galactic neutral hydrogen column density from Stark et al. (1992).

\end{table*}

\section {X-ray observations of M81 prior to $BeppoSAX$}

M81 has been pointed by many X-ray satellites. It was first observed
 in X-rays with {\it Einstein} (Elvis \& Van Speybroeck 1982;
 F88). This imaging satellite revealed the presence of several
 discrete sources in the M81 region, the brightest of which coincides
 in position with the nucleus (Fig. 1).  Its spectral
 representation was given in terms of a steep power law ($\Gamma=
 4.0$) or in terms of a thermal component of $kT= 1.1$ keV, both
 absorbed by a hydrogen column density $N_{\rm H}$ in excess of the
 Galactic value $N_{\rm H,Gal}$ (in Table 1). Subsequent observations by
 {\it GINGA} (Ohashi et al.  1992), {\it BBXRT} (Petre et al. 1993)
 and {\it ROSAT} (Boller et al. 1992) gave a spectral interpretation
 in terms of a power law with $\Gamma= 2$ and $N_{\rm H}$ ranging from
 being consistent with $N_{\rm H,Gal}$ to ten times higher.  Petre et
 al. also found a thermal component of $kT= 0.4$ keV.

{\it ASCA} pointed M81 several times between 1993 and 1999 (I96,
Iyomoto 1999). It was found that, over such a long period, the 2--10
keV luminosity of the nucleus was highly variable (Fig. 2).  However no
spectral variability was found, and the average spectrum was well
fitted by a power law continuum of photon index $\Gamma\sim 1.85$,
absorbed by a column density of $N_{\rm H}\sim 10^{21}$ cm$^{-2}$, plus a
thermal component with a temperature of $kT\sim 0.6-0.8$ keV. A
possibly broad or complex iron emission line centered at 6.6--6.9
keV, with an equivalent width of $170^{+60}_{-60}$ eV, was also
detected.

\section {X-ray data analysis}

M81 was observed by three of the Narrow Field Instruments onboard the
{\it BeppoSAX} satellite (Boella et al. 1997a): the Low Energy
Concentrator Spectrometer (LECS), the Medium Energy Concentrator
Spectrometer (MECS), and the Phoswich Detector system (PDS; Frontera
et al. 1997).  The journal of the observation is given in Table 2.
The LECS and MECS are grazing incidence telescopes with position
sensitive gas scintillation proportional counters in their focal
planes. The MECS, which at the epoch of the observation consisted of
two equal units, has a field of view of $56^{\prime}$ diameter, and
works in the range 1.6--10 keV (Boella et al. 1997b). The LECS
operates at softer energies (0.1--4.5 keV), has a field of view of
$37^{\prime}$ diameter, an energy resolution a factor of $\sim 2.4$
better than that of the {\it ROSAT} PSPC, but an effective area much
lower (between a factor of 6 and 2 lower, going from 0.3 to 1.5 keV;
Parmar et al. 1997).  The PDS is a collimated instrument, operating in
rocking mode (i.e., half of the time on the source and half on the
background direction), that covers the 13--300 keV energy band. It has
a triangular response with FWHM of $\sim 1^{\circ}\hskip -0.1truecm .3
$ (Frontera et al. 1997).

The cleaned and linearized data have been retrieved from the {\it
BeppoSAX} Science Data Center archive, and later reduced and analysed
using the standard software (XSELECT v1.4, FTOOLS v4.2, IRAF-PROS
v2.5, and XSPEC v10.0). For the MECS, we used the event file made by
merging the data of the two properly equalized MECS units.  The PDS
data reduction was performed using independently both XAS (v.2.0,
Chiappetti \& Dal Fiume 1997) and SAXDAS (v.1.3.0, Fiore et al. 1999)
software packages, and yielded consistent results.  In the following,
we will refer to the results obtained with the SAXDAS package.

\begin{table*}
\caption[Table 2.]{ Observation Log}
\begin{flushleft}
\begin{tabular}{lllllllllllll}
\noalign{\smallskip}
\hline
\noalign{\smallskip}
 Date        & \multicolumn{3}{l}{Exposure time$^a$ (ks)}  &\  & 
\multicolumn{3}{l}{Count Rate$^b$ (ct/s)}                 \\
\cline{2-4}\cline{6-8}
       & LECS & MECS & PDS & & LECS & MECS & PDS   \\
\noalign{\smallskip}
\hline
\noalign{\smallskip}
1998 Jun 4  & 43.20 & 100.3 & 36.40 & & 0.329$\pm 0.003$ & 0.496$\pm 0.002$ & 
0.57$\pm 0.04$     \\ 
\noalign{\smallskip}
\hline
\end{tabular} 
\end{flushleft}
\bigskip

$^a$ On-source net exposure time. The LECS exposure time is
considerably shorter than the MECS one, because the LECS
can operate only when the spacecraft is not illuminated by the Sun.

$^b$ Background subtracted source count rates, with photon counting
statistics errors.

\end{table*}

\subsection{Spatial analysis and extraction regions}

The satellite pointed the optical center of the galaxy (in Table 1).
The center of the X-ray emission was found to be located at $9^h 55^m
26^s\hskip -0.1truecm .5$, $69^{\circ} 03^{\prime} 43 ^{\prime\prime}$
for the LECS, and $9^h 55^m 32^s \hskip -0.1truecm .5$, $69^{\circ}
03^ {\prime} 35^{\prime\prime}$ for the MECS, i.e., within $\sim
1^{\prime}$ from the optical center of the galaxy given in Table
1. Therefore, within the accuracy with which positions are given by
$BeppoSAX$, the X-ray centers coincide with the M81 nucleus.  The LECS
and MECS images in the 0.1--4.0 and 1.65--10.5 keV energy bands
respectively are shown in Fig. 1.  Also shown are the positions of the
sources detected in the field by the {\it Einstein} observation.

\begin{figure*}
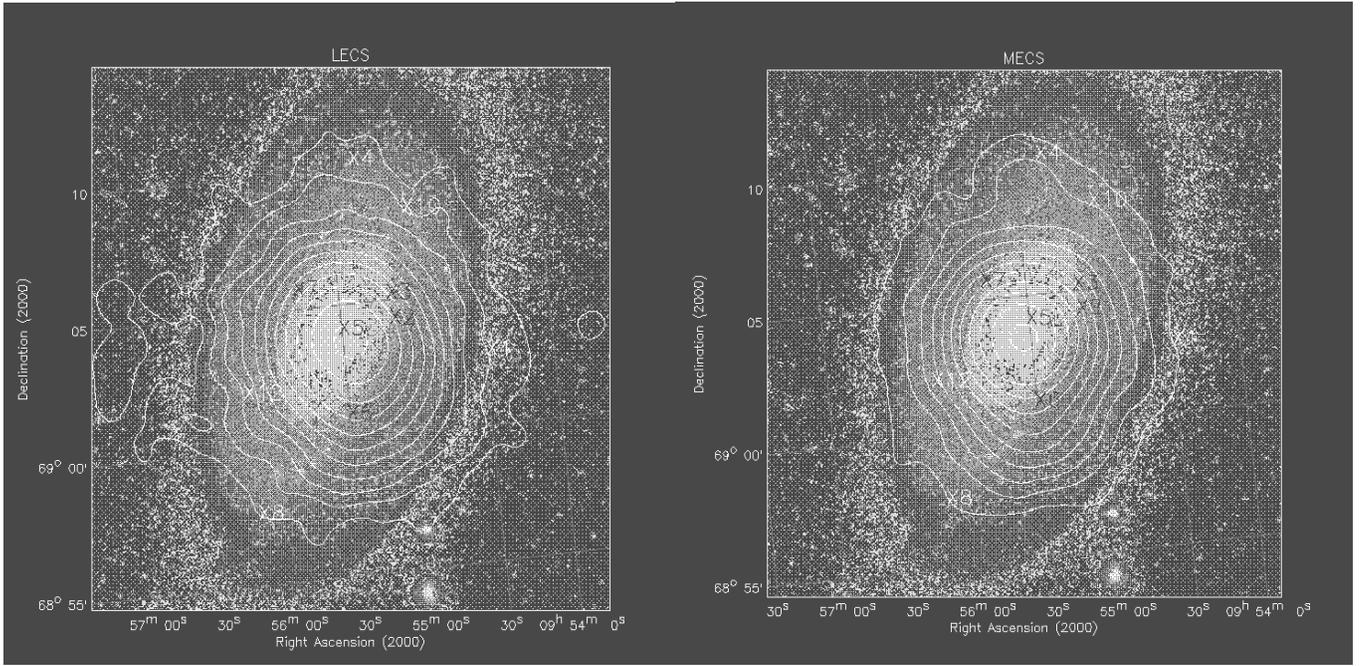

\parbox{10cm}{
\psfig{file=1563.f1a,width=9cm,height=8.8cm,angle=0} }
\ \hspace{-0.5cm} \
\parbox{10cm}{
\psfig{file=1563.f1b,width=9cm,height=8.8cm,angle=0} }
\caption[]{The LECS and MECS images of M81, smoothed with a gaussian of $\sigma
=32^{\prime\prime}$. Contours plotted are logarithmically spaced from 0.7
to 90\% of the peak intensity, for the LECS, and from 0.6
to 90\% for the MECS.
The labels X1--X12 mark the positions of the sources detected by {\it Einstein}
in the field (F88); X5 is the nucleus of M81. The optical image of M81 is from the Digital Sky Survey of
the Space Telescope Science Institute.}
\end{figure*}

For a point source, the PSF of the MECS includes 80\% of photons of
energies $\geq 1.5$ keV within a radius of $2^{\prime}\hskip
-0.1truecm .7$ (Boella et al. 1997b). The PSF of the LECS is broader
than that of the MECS below 1 keV, while it is similar to it above 2
keV (see http://www.sdc.asi.it/software/cookbook).  In order to study
the spectral properties of the nucleus of M81, given these
instrumental characteristics, counts have been extracted from a circle
of $3^{\prime}$ radius for the MECS, and of $4^{\prime}$ radius for
the LECS. The larger radius adopted for the LECS is motivated by its
larger PSF at softer energies. A radius of $3^{\prime}$ was also
adopted for the extraction of the {\it Einstein} and {\it ASCA}
spectra (F88, I96). We estimate in Sect. 3.6 the contribution of the
unresolved emission from the galaxy to the LECS and MECS spectra
extracted as described above.

The starburst galaxy M82, that lies $\sim$ 37$^{\prime}$ from M81, is
inside the PDS field of view (and outside the LECS and MECS
ones). Given the PDS triangular response ($1^{\circ}\hskip -0.1truecm
.3$ FWHM), half of the M82 flux between 13--300 keV could
``contaminate'' the M81 observation.  A $BeppoSAX$ observation of M82
performed one year earlier indicates, however, that M82 has a steeper
spectrum than M81 (photon index of $3.8\pm 1.8$ in the PDS energy
band), and a 13--50 keV flux $\lsim 1.7\times 10^{-11}$ erg cm$^{-2}$
s$^{-1}$ (Cappi et al. 1999), i.e., a factor of $\gsim 2$ lower than
that of M81 in the same energy range.  We estimate then that the PDS
spectrum of M81 could be contaminated by at most 25\% of its
flux. Therefore, we added a systematic uncertainty of this level to
the spectral data between 13--50 keV.

\subsection {Variability}

\begin{figure}[htb]
\parbox{10cm}{
\psfig{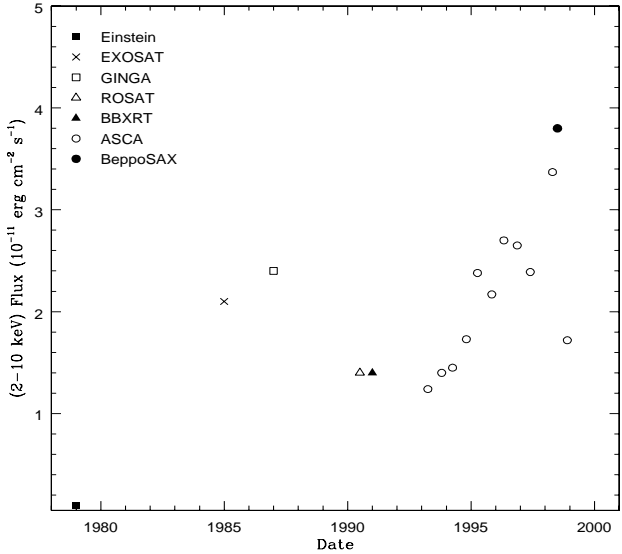} }
\ \hspace{0cm} \
\caption[]{ Long term variation of the (2--10) keV flux from the
nucleus of M81.  The points refer to observations by
{\it Einstein, EXOSAT, GINGA, ROSAT, BBXRT, ASCA}, and {\it BeppoSAX}. Typical
90\% confidence errors on fluxes are not larger than $\pm$10\%. The
fluxes coming from $ASCA$ data have been kindly provided by N. Iyomoto
(1999), and that from $EXOSAT$ data by P. Barr (priv. comm.). 
For the (2--10) keV fluxes of {\it Einstein} and $ROSAT$
we have adopted the extrapolations by Petre et al. (1993).}
\end{figure}

The long term variation of the M81 (2--10) keV flux, obtained by
collecting observations from various observatories, is plotted in
Fig. 2. During the {\it BeppoSAX} pointing the nucleus was at the
highest level ever observed. From 1993 to 1998 it brightened by a
factor of $\sim 4$.

A short term variation is clearly detected by $BeppoSAX$; the light
curves obtained from the LECS and MECS data are plotted in Fig. 3. The
count rates of both instruments show a variation of $\sim 30$\% from
peak to valley over roughly one day.  The fit of the MECS light curve
with the sine function (Fig. 3, bottom) is significantly better than
that with a constant value; the period turns out to be almost 2 days
(43 hours). Since this period is comparable to the length of the
observing time, the presence of a periodical change cannot be
established; future observations might reveal whether this periodicity
is real.  The $BeppoSAX$ light curve extends over a time interval as
long as the longest of the {\it ASCA} observations (made during April
1993; Serlemitsos et al. 1996). The {\it ASCA} light curve in that
case showed variability of the order of $\sim 20$\% on a timescale of
$\sim 1$ day.

\begin{figure}[htb]
\parbox{10cm}{
\psfig{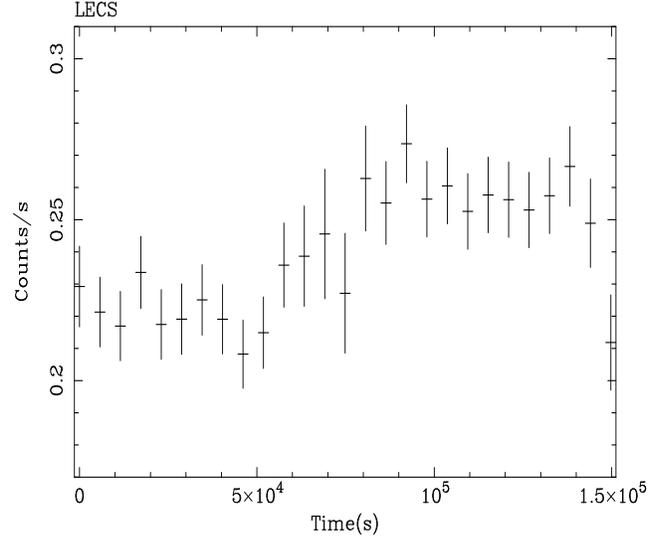} }
\ \hspace{0cm} \

\parbox{10cm}{
\psfig{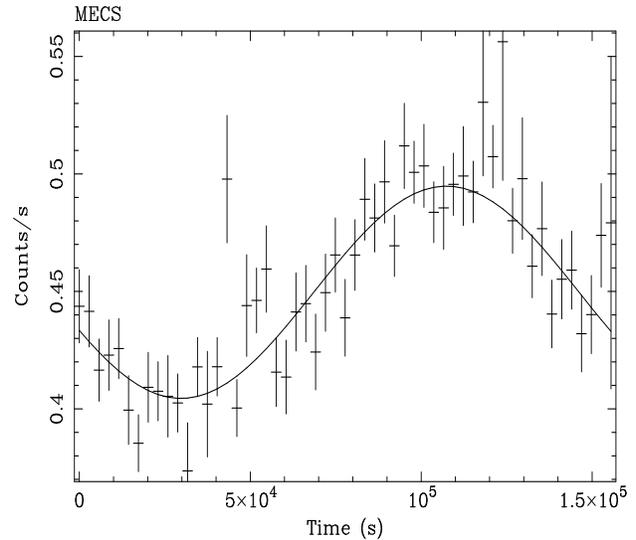} }
\ \hspace{0cm} \
\caption[]{{\it Top}: LECS light curve; counts are from $4^{\prime} $ radius, and 0.1--4 
keV. {\it Bottom}: MECS light curve; counts are from $3^{\prime} $ radius, and 2--10
 keV. The sine function that best fits the data is also plotted. }
\end{figure}

It is unlikely that the flux variation is produced by an X-ray binary
in the field, because it should be extremely luminous ($L_{\rm X}\sim
2\times 10^{40}$ erg s$^{-1}$ in the 0.1--2 keV band, and $L_{\rm
X}\sim 10^{40}$ erg s$^{-1}$ over 2--10 keV).  No sources so luminous
have been detected by {\it Einstein} and $ROSAT$ (see Sect. 3.6).
Another fact points against the hypothesis of a binary responsible for
the variability: the variation detected by {\it ASCA}, which had
observed the nucleus of M81 during a phase of lower flux level in 1993
(Fig. 2), had a smaller amplitude than that detected by {\it
BeppoSAX}; a variation of larger amplitude is instead expected when
the flux of the nucleus is lower, if it is produced by a binary.

\subsection {Spectral analysis}

 The background spectrum was estimated from blank fields event files
(released in Nov. 1998).  These blank-fields events were accumulated
on five different pointings of empty fields; the extraction regions
used for them correspond in size and position to those of the source.
Spectral channels corresponding to energies 0.12--4 keV, 1.65--10.5
keV, and 15--200 keV have been used for the analysis of
the LECS, MECS and PDS data respectively.

The data have been compared to models, convolved with the instrumental
and mirror responses, using the $\chi ^2$ minimization method. For
this comparison the original channels have been rebinned in order to
sample the instrument resolution with the same number of channels
at all energies, and in order to have bins adequately filled for
applicability of the $\chi^2$ statistic to assess the goodness of 
fit (Cash 1979).  The spectral response matrices and effective area
files released in September 1997 have been used in the fitting
process.  We fitted the models simultaneously to the LECS, MECS and
PDS data.  In the fitting two normalization constants have been
introduced to allow for known differences in the absolute
cross-calibrations between the detectors (Fiore et al. 1999).  The
models are modified by photoelectric absorption of the X-ray photons
due to intervening cold gas along the line of sight, of column density
$N_{\rm H}$. The results of the spectral analysis are presented in Table 3.

\begin{table}
\caption[]{ Spectral analysis}
\begin{flushleft}
\begin{tabular}{  l    l }
\noalign{\smallskip}
\hline 
\noalign{\smallskip}
 Model parameters    &  Best fit values  \\
\noalign{\smallskip}
\hline 
\noalign{\smallskip}
\hline
\noalign{\smallskip}
A. pow+gauss:&            \\
\noalign{\smallskip}
$10^{-20}N_{\rm H}^a$ (cm$^{-2}$) &  7.2 (6.1--8.5)    \\
$\Gamma $     &  1.86 (1.84--1.89)  \\
$10^{11}$F(erg cm$^{-2}$ s) &  2.1, 3.8, 7.4     \\
Line E(keV)   &  6.70 (6.59--6.81) \\
Line EW (eV)  &   104 (64--133)    \\
$\chi^2/\nu$  &   193/178     \\
\hline
\noalign{\smallskip}
B. mekal$^b$+pow:&             \\
\noalign{\smallskip}
$Z$ ($Z_{\odot}$)    &  0.5 (0.4--1.3)  \\
$kT$ (keV)        & 6.5 (3.5--7.9)   \\
$10^{11}$F$_{mekal}$(erg cm$^{-2}$ s) &  0.4, 0.7, 0.6    \\
\noalign{\smallskip}
$10^{-20}N_{\rm H}^a$ (cm$^{-2}$)&  7.1 (5.7--8.2)  \\
$\Gamma $    & 1.85 (1.76--1.88) \\
$10^{11}$F$_{pow}$(erg cm$^{-2}$ s) &  1.7, 3.1, 5.6  \\
$\chi^2/\nu$ & 181/177  \\
\hline
\noalign{\smallskip}
C. pow+gauss+edge:&            \\
\noalign{\smallskip}
$10^{-20}N_{\rm H}^a$ (cm$^{-2}$)&  6.7 (5.7--7.9)   \\
$\Gamma $     &  1.84 (1.82--1.87)  \\
$10^{11}$F(erg cm$^{-2}$ s) &  2.1, 3.8, 7.4     \\
Line E(keV)   &  6.69 (6.58--6.83) \\
Line EW (eV)  & 84 (48--121)    \\
Edge E(keV)   & 8.6 (7.8--9.0) \\
$\tau $       & 0.15 (0.07--0.24) \\
$\chi^2/\nu$  &   182/176     \\
\hline
\noalign{\smallskip}
D. mekal$^b$+pow+gauss+edge:&            \\
\noalign{\smallskip}
$Z$ ($Z_{\odot}$)    &  0.03 (0.--0.09)  \\
$kT$ (keV)        & 0.50 (0.25--0.95)   \\
$10^{11}$F$_{mekal}$(erg cm$^{-2}$ s) &  0.3, 0, 0    \\
\noalign{\smallskip}
$10^{-20}N_{\rm H}^a$ (cm$^{-2}$)&  12.0 (9.0--15.6) \\
$\Gamma $     &  1.86 (1.83--1.89)  \\
$10^{11}$F$_{pow}$(erg cm$^{-2}$ s) &  1.8, 3.8, 7.2     \\
Line E(keV)   &  6.70 (6.58--6.83) \\
Line EW (eV)  & 95 (49--123)    \\
Edge E(keV)   & 8.6 (7.8--9.1) \\
$\tau $       & 0.14 (0.06--0.23) \\
$\chi^2/\nu$  &   172/173    \\
\hline
\hline
\end{tabular} 
\end{flushleft}

\smallskip
$^a$ Column density of neutral hydrogen in addition to $N_{\rm H,Gal}$.

$^b$ The column density of neutral hydrogen absorbing the mekal component 
is fixed at the Galactic value of $4.1\times 10^{20}$cm$^{-2}$.

\smallskip

Note: $\nu $ is the number of degrees of freedom of the fit. The
values between parenthesis, close to the best fit values, give the
90\% confidence intervals for one interesting parameter (these
intervals correspond to $\Delta \chi^2=2.71$).

\smallskip   

Note: three values for the absorbed fluxes $F$ are given, respectively for the 
(0.1--2), (2--10), and (10--100) keV energy bands. 

\end{table}

\subsubsection{Fitting results}

A single thermal model cannot reproduce the broad band data, in which
the nucleus is detected up to 100 keV. The LECS and MECS data
constrain the temperature to be around 8 keV, but with this
temperature the model of course totally fails to reproduce the PDS
data. On the other hand, a thermal component that fits the high energy
region would have $kT\sim 90 $ keV, and would be too flat to reproduce
the LECS and MECS data.  A two-component thermal model (with
temperatures $kT\sim 3$ and $kT\sim 60$ keV) gives a better fit, but
still of poor quality.  A single power law component of photon index
$\Gamma \sim 1.85$ can instead reproduce the data over the whole
energy band (Fig. 4).  As can be seen looking at the residuals (see
also Fig. 5), excess emission is present between 6--7 keV, as already
found by $ASCA$ (Sect. 2). Adding a narrow gaussian line to the
model gives a decrease of $\chi^2$ of 21 for two additional parameters
(the line center energy and normalization); so, this line is
statistically significant at $>99.9$\% confidence level in an F-test
(this follows the treatment, specific for 
fitting of X-ray spectral data,  about how to assess
the usefulness of adding extra parameters to an initial model 
by Malina et al. 1976).  The
line is centered at $6.70^{+0.11}_{-0.11}$ keV, and its equivalent
width is EW=$104^{+29}_{-40}$ eV (see model A in Table 3; Fig. 7). If
the width $\sigma $ of the line is left as a free parameter, we obtain
for it at the best fit a value of $110 $ eV and an upper limit at 90\%
confidence for one interesting parameter of 300 eV. The 6.4 keV value
for the line center energy is excluded at $>99$\% confidence level.
Its introduction in the fitting provides a 90\% confidence upper limit
on its EW of 42 eV, and no improvement of the fit.

\begin{figure*}
{\psfig{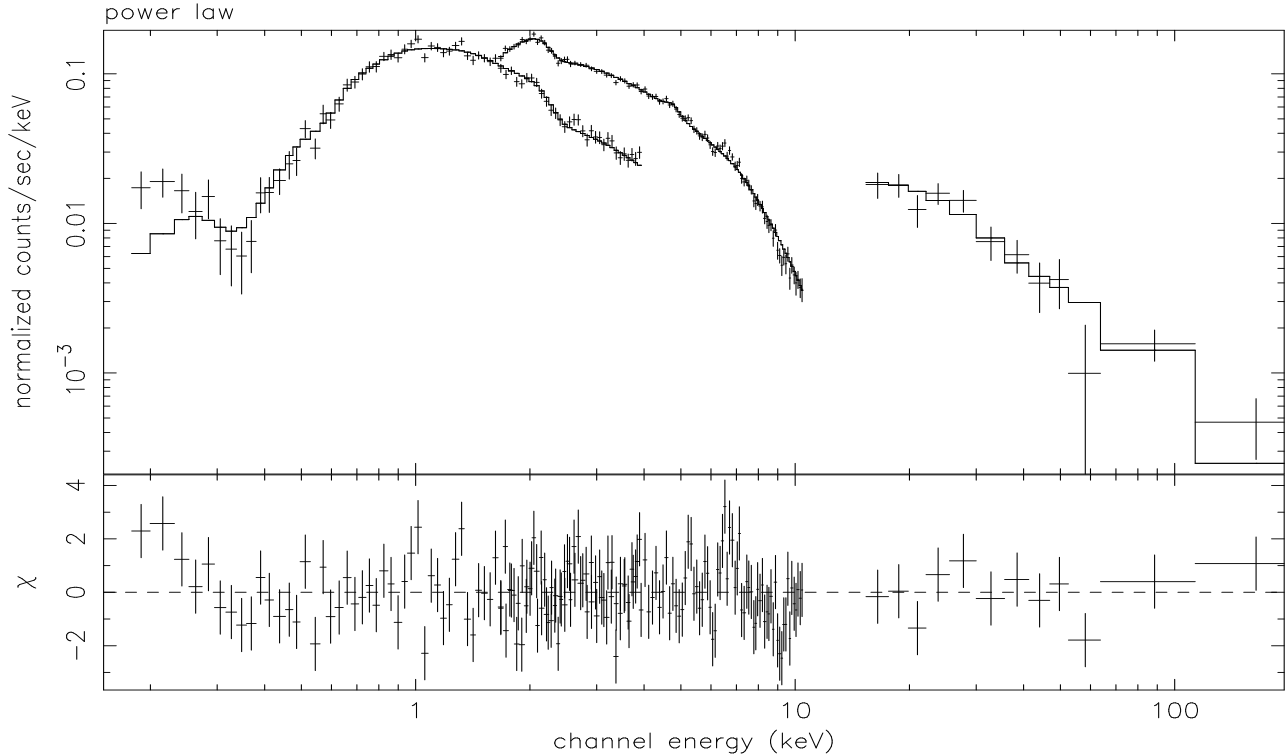} }
\ \hspace{0cm} \
\caption[]{ {\it BeppoSAX} LECS, MECS and PDS observed spectra of M81 
(crosses), 
 modeled with a power law of $\Gamma=1.85$ (solid line). The residuals, in terms of $\sigma$'s, between the 
data and the model are plotted below.  }
\end{figure*}

\begin{figure}
\parbox{7cm}{
\psfig{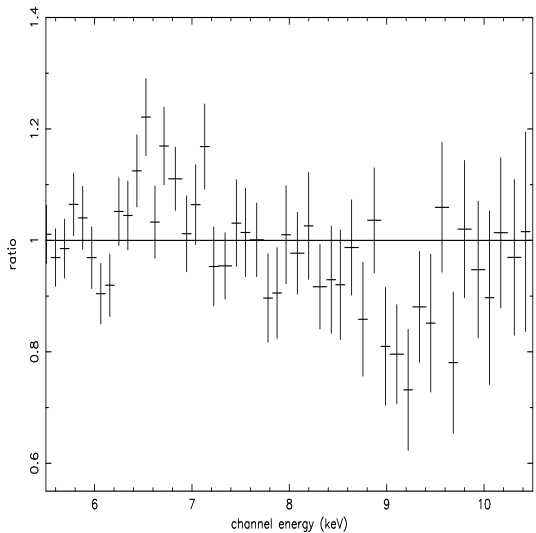} }
\ \hspace{0cm} \
\caption[]{ The data/model ratio for the best fit continuum model
(the power law shown in Fig. 4) around the 6.7 keV emission line and the
8.6 keV absorption edge.}
\end{figure}

The power law is absorbed by cold material exceeding the Galactic
column density by $N_{\rm H}\sim 7\times 10^{20}$ cm$^{-2}$.  This intrinsic
column density  is consistent with that
derived optically: Filippenko \& Sargent (1988) estimated an
E(B--V)=0.1$^{+0.15}_{-0.1}$ for the broad line region of M81, which
translates into $N_{\rm H}=5.8\times 10^{21} E(B-V)=5.8\times 10^{20}$
cm$^{-2}$.

A further improvement in the quality of the fit is obtained by fitting
with a power law plus a thermal component (model B in Table 3; the
mekal model describes the thermal emission from an optically thin hot
plasma, both from continuum and lines). Then  $\Gamma \sim 1.85$ and again
$N_{\rm H}\sim 10^{21}$ cm$^{-2}$ for the power law component; for the
thermal component $kT\sim 6$ keV, $N_{\rm H}\sim N_{\rm H,Gal}$ (imposed,
otherwise it would be lower) and $Z\sim 0.5 Z_{\odot}$.  This model
gives a fit of better quality than model A because the thermal
component reproduces the 6.7 keV line and gives slightly smaller
residuals than a simple power law in a few channels below 1 keV.  This
 thermal component contributes 16\% of the total unabsorbed flux
over 0.1--10 keV.

\begin{figure*}[htb]
{\psfig{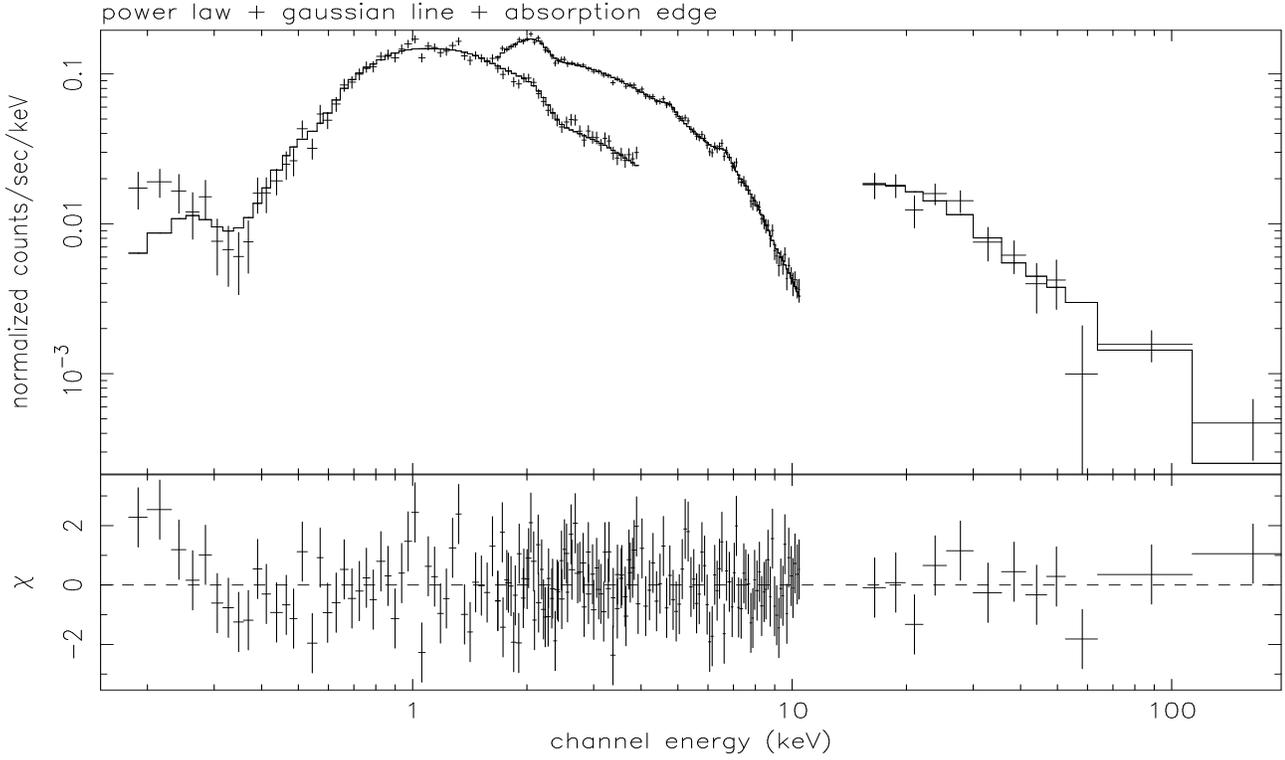} }
\ \hspace{0cm} \
\caption[]{ {\it BeppoSAX} LECS, MECS and PDS observed spectra of M81
(crosses), modeled with a power law plus gaussian line plus absorption
edge (model C in Table 3).  The residuals, in terms of $\sigma$'s,
between the data and the model are plotted below.  }
\end{figure*}

\subsubsection {Detection of an ionized absorber}

There is another spectral region, located around 9 keV, that shows
 residuals when fitted with a simple power law (Figs. 4 and 5).  In
 fact the quality of the ``power law plus gaussian line" fit improves
 by adding an absorption edge (Fig. 6; model C in Table 3).  This
 turns out to be located at $E= 8.6^{+0.4}_{-0.8}$ keV, with optical
 depth $\tau= 0.15 ^{+0.09}_{-0.08}$ (Fig. 8).  The presence of one
 edge is statistically significant at $>99$\% confidence level (from
 an F-test; $\chi^2$ decreases by 11 for 2 more fitting
 parameters). We cannot exclude that the `hole' around 9 keV could be
 produced by more than one ionization stage.  Fitting with two edges,
 at energies fixed at 8.5 keV (FeXXIII) and 8.8 keV (FeXXV), improves
 the fit slightly but not significantly.  This is the first time an
 absorption edge is detected in the X-ray spectra of LINERs.

\begin{figure}
\parbox{7cm}{
\psfig{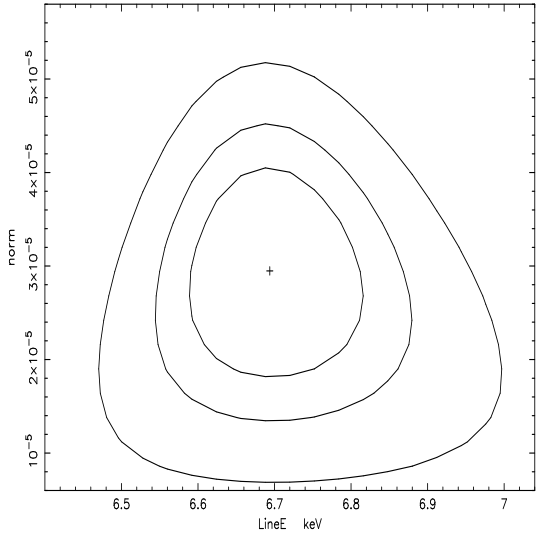} }
\ \hspace{0cm} \
\caption[]{ The 68\%, 90\% and 99\% confidence contours for two interesting
parameters for the line energy and normalization (in units of photons cm$^{-2}$
s$^{-1}$ in the line).}
\end{figure}

\begin{figure}[htb]
\parbox{7cm}{
\psfig{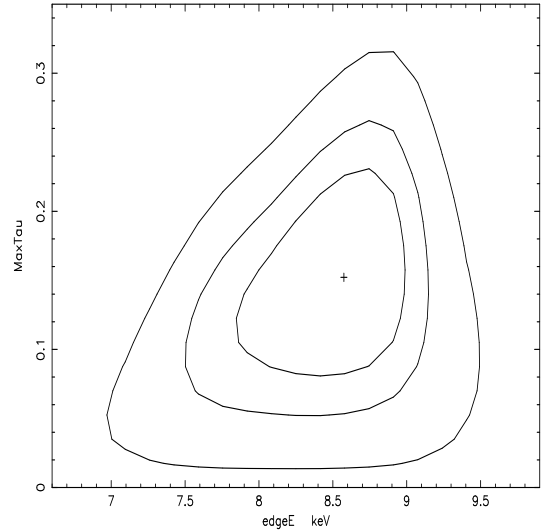} }
\ \hspace{0cm} \
\caption[]{ The 68\%, 90\% and 99\% confidence intervals for two interesting
parameters  for the energy and optical depth
$\tau$ of the absorption edge. }
\end{figure}

\subsection{Spectral analysis and variability}

We have also looked for spectral differences between the two states of
low and high flux, i.e., we have studied separately the spectra
obtained from the first and the last $\sim 7.5\times 10^4$ s of the
pointing (see Fig. 3).  We do not find significant changes between the
two states in any spectral property described above (power law slope,
emission line and absorption edge center energies, optical depth of
the edge). Going from the low to the high state the normalization of
the power law increases, and the equivalent width of the 6.7 keV line
marginally decreases.  More precisely, when a fit with a power law +
gaussian line + edge is performed, going from the low to the high
state $\Gamma $ slightly decreases from $1.86^{+0.03}_{-0.04}$ to
$1.84^{+0.03} _{-0.03}$, while the EW drops from $130^{+64}_{-42}$ eV
to $55^{+47}_{-47}$ eV. The line center energy goes from 6.73 to 6.65
keV, with the associated confidence intervals almost overlapping. The
line flux goes from $4.4\times 10^{-5}$ photons cm$^{-2}$ s$^{-1}$ to
$2.1\times 10^{-5}$ photons cm$^{-2}$ s$^{-1}$.

\subsection {X-ray fluxes and luminosities}

In Table 3 we report the observed fluxes in the energy bands (0.1--2),
(2--10) and (10--100) keV. The luminosities are obtained from fluxes
corrected for the observed total $N_{\rm H}$, adopting a distance of 3.6
Mpc. So, for the model power law + gaussian line + edge (model C), in
the three bands $L_{\rm X}=7.5,\,4.3,\,9.4\times 10^{40}$ erg s$^{-1}$.  The
(0.1--100) keV luminosity is $2.1\times 10^{41}$ erg s$^{-1}$.

\subsection {Background emission from the M81 galaxy}

Before ascribing the derived spectral properties to the
nucleus of M81, we must evaluate the contribution, within the region
used to extract the LECS and MECS spectra, from the various possible
non-nuclear X-ray components of M81: stars, supernova remnants, X-ray
binaries, and possibly a hot ISM.  {\it Einstein} resolved a few
sources to within $3^{\prime}$ (Fig. 1).  In particular, we expect a
contribution from sources X2, X3, X6, and X7.  Given their positions,
their {\it Einstein} fluxes (F88), and the shape of the PSF, we
estimate their contribution to be $< 6$\% of the total net counts
within a radius of $3^{\prime}$ from the nucleus\footnote{During the
{\it Einstein} pointing the source X5 was in a low state, by at least
a factor of ten lower than what was found later by $EXOSAT$, $GINGA$,
$ROSAT$, $BBXRT$, and a factor of $\sim $ fourty lower than found by
$BeppoSAX$ (Fig. 2); so, using the {\it Einstein} fluxes for the
various point sources we can estimate an upper limit to their
contribution to the total net counts.}.  The point sources resolved in
the {\it ROSAT} HRI image within $3^{\prime}$ of the nucleus are 11
(G. Fabbiano, private communication). Their HRI count rates give a
contribution of $\lsim 6$\% to the total (0.1--2) keV flux within a
circle of $3^{\prime}$ radius.  Note that there are no sources bright
enough to produce the short term variability shown by $BeppoSAX$
(Fig. 3, Sect. 3.2).  From an estimate kindly provided by
G. Fabbiano of the `galaxy background' contribution (which includes
also the $ROSAT$ unresolved emission) we derive that $\gsim 85$\% of
the total counts within $3^ {\prime}$ belong to the active nucleus,
in the $ROSAT$ (0.1--2) keV band; this calculation takes into account
also the increase in flux observed between the $ROSAT$ and $BeppoSAX$
pointings (Fig. 2).

The galaxy background contribution is difficult to estimate with data
from satellites sensitive to higher energies than $ROSAT$, due to
their poor angular resolution. This contribution has proven to be not
influential for the spectral analysis in the {\it ASCA} (0.5--10) keV
band (I96).  I96 estimated that the galaxy background outside the
central $3^{\prime}$ is well fitted by a bremsstrahlung of $kT\sim 8$
keV (which led I96 to ascribe the galaxy background to the integrated
emission from LMXBs). A spectral fit of the nuclear region with such a
bremsstrahlung component restricted the galaxy background contribution
to $<12$\% of the 2--10 keV luminosity of the active nucleus. A
similar analysis for the LECS and MECS data would not give results
more reliable than those obtained with $ASCA$, due to the angular
resolution of these instruments and the vignetting effect that becomes
large off-axis.  All in all, we are confident that $> 80$\% of the
(0.1--100) keV net counts extracted from a circle of $3^{\prime}$
radius from the galaxy center come from the active nucleus.
 
\section { Discussion}

The simplest model that explains all the features of the 0.1--100 keV
spectrum is a power law of $\Gamma\sim 1.84$, to which a gaussian line
at 6.7 keV and an absorption edge at $\sim 8.6$ keV are added.  The
energy of the line indicates K$\alpha$ emission from He-like iron
(FeXXV), while its 90\% confidence interval goes from FeXXII to
FeXXV. The absorption edge can be produced by ions from FeXVII to
FeXXV, within the 90\% confidence interval, and its best fit energy
corresponds to the FeXXIV K-edge ($\sim 8.6$ keV; Makishima
1985). Both features therefore come from highly ionized
material. Moreover they could be consistent with each other also in
column density.  The best fit value for the optical depth $\tau$, when
assuming a cosmic abundance of iron, and the photoionization cross section
for the K-shell for ionization stages from FeXXIII to FeXXV (Krolik \&
Kallman 1987), corresponds to a hydrogen column density of $\sim 2\times
10^{23}$ cm$^{-2}$ for the ionized material.  The observed EW of $\sim
100$ eV for the Fe-K line can be produced by reflection from or
transmission through a column density of $N_{\rm H}\sim 10^{23}$ cm$^{-2}$,
in the case of neutral material (Makishima 1985). In case of highly
ionized material, an EW of $\sim 100$ eV is expected to correspond to
slightly lower column densities than for the neutral case (Turner et
al. 1992). 

As can be seen from Table 3, a model made by the superposition of a
power law and a thermal component at $kT\sim 6$ keV (model B) is
equally statistically acceptable as a model with a power law +
gaussian line + edge (model C), because it reproduces the 6.7 keV
line, and the residuals in a few channels below 1 keV are slightly
smaller.  Since the absorption edge remains totally unexplained with a
thermal model, we prefer to adopt model C.  To reduce its residuals
below 1 keV, the addition of a soft ($kT\sim 0.5$ keV) thermal
component proved effective (model D in Table 3); it contributes $\sim
12 $\% of the (0.1--2) keV absorbed flux.  This soft component is not
required by the data though, since $\chi^2$ is reduced by ten, but the
number of free parameters increases by three.
A soft component had been detected also in the analysis of $BBXRT$
data by Petre et al. (1993), and by I96 (Sect. 2). This soft thermal
emission could represent the `galaxy background' within $3^{\prime}$
of the nucleus, that we estimate to account for $\lsim 15$\% of the
total emission within that radius in the $ROSAT$ band (Sect.
3.6). Soft emission with this temperature and sub-solar abundance is
common to spiral galaxies, and may well come from heterogeneous
origins (supernova remnants, stars, thermal gas created in a
starburst; e.g., Serlemitsos et al. 1996, Iyomoto et al. 1997).

The possible origin of the adopted model (power law + gaussian line +
edge) is discussed in the following.  The power law component that
well reproduces the (0.1--100) keV spectrum is usually related to the
presence of an AGN. This presence is not unexpected here: based on the
velocities and dimensions of the broad line region, and assuming that
gravity dominates the motion of the clouds, Ho et al. (1996) derived a
central mass of $(0.7-3)\times 10^6 M_{\odot}$ for the nucleus of M81,
recently updated to $4\times 10^6 M_{\odot}$ (Ho 1999).  Moreover, no
absorption much in excess of the Galactic value is seen in the X-ray
spectrum.  This property, together with the value of $\Gamma$ close to
that found for Seyfert 1's (Nandra \& Pounds 1994), makes M81 similar
to Seyfert 1's. 

\subsection {Origin of the iron-K emission and absorption}

Are the 6.7 keV emission line and the 8.6 keV absorption edge produced
by reflection (e.g., from an X-ray illuminated disk, $\dot {\rm
Z}$ycki \& Czerny 1994), or by transmission?  And where is the highly
ionized material responsible for the line and the edge located?  In
the case of reflection from a disk, its ionization state should be
higher than in Seyfert 1's, because the iron line energy is higher
(6.7 keV instead of 6.4 keV).  But there are both observational and
physical problems with this hypothesis. The observational problem is
that the expected reflection continuum from highly ionized matter is
absent in the spectral data. The fit  with the inclusion of
a reflection component from an ionized face-on disk shows that this
component is not required by the data, and we derive a 90\% confidence
upper limit on the subtended solid angle $\Omega/2\pi$ of the
reflecting material of 0.3. Moreover, the
fit is of poor quality because reflection cannot reproduce a sharp
absorption edge at 8.6 keV as observed (it produces smeared edges,
Ross et al. 1999). The physical problem is that the observed accretion
rate is too low for the required ionization level in the disk.  In
order to have He-like iron the accretion rate in Eddington units $\dot
m=\dot M/\dot M_{\rm Edd}$ must be $\sim 0.4$ ($\geq 0.2$ for $\geq
$FeXXII; Matt et al. 1993). The observed bolometric luminosity of the
nucleus gives $\dot m\sim 4\times 10^{-4}$ only (see Sect. 4.2
below). So, $\dot m$ fails severely the requirement $\dot m\geq
0.2$. Therefore the line and the edge are unlikely to be produced by
reflection from a highly ionized accretion disk.

We suggest instead that they are produced by transmission through a
 highly photoionized medium, located close to the nucleus.  If so,
 this points out an interesting similarity with the presence of a warm
 absorber in Seyfert 1's (Nandra \& Pounds 1994).  Usually Seyfert 1's
 show the oxygen absorption edges, superimposed onto the power law
 continuum, at 0.74 and 0.87 keV; we do not observe these edges,
 because the material is highly ionized. Concerning the location of
 such material, it could be close to the source of ionizing
 photons. In the hypothesis that its line of sight thickness is much
 less than $d$, its distance from the source of photoionizing photons,
 Reynolds \& Fabian (1995) derive a constraint on $d$: $d<L/N\xi$,
 where $L$ is the luminosity of ionizing photons, $N$ is the column
 density of photoionized matter, and $\xi$ is its level of
 ionization\footnote{ In order to have an ionization level $\geq
 $FeXXII to peak, the ionization parameter $\xi \gsim 100$ erg cm
 s$^{-1}$, while $\xi\sim 120$ erg cm s$^{-1}$ for He-like iron to
 peak. These values refer to an input spectrum of a power law with
 $\Gamma =1.5$ from 13.6 eV to 13.6 keV, passing through an optically
 thin shell of gas (Turner et al. 1992).} ($\xi=L/nd^2$, where $n$ is
 the number density of photoionized matter, Kallman \& McCray 1982).
 So $d< 10^{16}$ cm, if $N\sim 2\times 10^{23}$ cm$^{-2}$ (as
 derived from the absorption edge in Sect. 4) and  $L\sim 2
 \times 10^{41}$ erg s$^{-1}$ for the observed (0.1--100) keV
 luminosity of the nucleus of M81. This places the
 absorber slightly closer to the nucleus with respect to what is found
 typically for Seyfert 1's, where the so-called warm absorber is
 located at radii coincident with, or just outside, the broad line
 region (e.g., Reynolds \& Fabian 1995).

\subsection {Origin of the X-ray continuum}

In the framework of standard $\alpha$-disks (Shakura \& Sunyaev 1973)
X-ray emission with a power law shape arises from inverse Compton
scattering, by a hot optically thin plasma in a corona, of UV/soft
X-ray photons produced by the accretion disk (see, e.g., Svensson 1996
for a review).  The observed bolometric luminosity of the nucleus of
M81 corresponds to a low accretion rate.  Adopting the central mass
value of $4\times 10^6 M_{\odot}$, the bolometric luminosity of the
nucleus obtained by integrating the observed spectral energy
distribution from the radio to 10 keV ($L=2.1\times 10^{41}$ erg
s$^{-1}$, Ho 1999) and a radiative efficiency of 0.1, it turns out
that $\dot m\sim 4\times 10^{-4}$.  A problem with this scenario is
that we do not detect signs of reflection from optically thick cold
material, usually found in Seyfert 1's, and attributed to the presence
of an accretion disk. These signs are a 6.4 keV emission line with
typical EW of 100--150 eV, and a broad bump peaking at 10--20 keV
(Nandra \& Pounds 1994). As a further comparison with the continuum of
classical, more luminous AGNs, note that also the `big blue bump',
traditionally attributed to thermal radiation from an accretion disk,
is absent in the spectral energy distribution observed for the nucleus
of M81 (Ho 1999).  So, we conclude that this LLAGN is not a simple
extension of high luminosity ones.

An alternative possibility for the origin of the continuum emission
 could be the presence of an advection dominated accretion flow (ADAF,
 Narayan \& Yi 1995), a solution devised for AGNs radiating at very
 low Eddington ratios (e.g., for NGC4258, Lasota et al. 1996).  In
 ADAF models at low mass accretion rates ($\dot m \lsim 10^{-3}$) the
 X-ray emission has a significant thermal bremsstrahlung contribution
 produced by the electrons in the flow, at $kT\sim 10^9-10^{10}$ K;
 this corresponds to a very flat power law ($\Gamma \lsim 1.3$) up to
 several tens of keV. As $\dot m $ gradually approaches and exceeds
 $10^{-3}$ the ADAF model predicts that inverse Compton scattering of soft
 synchrotron photons by the flow electrons will dominate 
 the X-ray emission more and more. This second regime of the model is
 required to be at work for M81, in order to explain the steep
 spectral shape revealed by $BeppoSAX$ (Sect. 3.3.1; this is
 possible when assuming a radiative efficiency $<0.1$).

In conclusion, the origin of the power law emission is an open question.
 From X-ray, UV and optical observations there is no evidence for an
 accretion disk, and the conditions for an ADAF are satisfied, but the
 slope of the X-ray continuum is very similar to that of bright
 Seyfert galaxies, and it remains to be investigated whether it can be
 reproduced by an ADAF.

\subsection {Comparison with {\it ASCA} results on LINERs}

How common are the results presented here among LINERs?  On the basis
of $ASCA$ data, a canonical model was found to work for LLAGNs, LINERs
and starburst galaxies by Ptak et al. (1999). This model consists of a
soft component of $kT\sim 0.5-1 $ keV, usually spatially extended and
with absorption consistent with the Galactic value, plus a power law
of mean $\Gamma \sim 1.7$, with small or no intrinsic absorption, {\it
or} another thermal component of $kT\sim 6$ keV.  This model turned
out to be successful also in the case of the detailed study of the
LINERs NGC4579 (Terashima et al. 1998) and NGC4736 (Roberts et
al. 1999). The results obtained here from $BeppoSAX$ data are in
agreement with this canonical model, with the addition that we can
firmly establish that a power law is to be preferred to the hard
thermal component.  The presence of a 6.7 keV gaussian line, and no
evidence of a 6.4 keV line, was also found from $ASCA$ data for the
LINERs NGC4579 (Terashima et al. 1998) and NGC4736 (Roberts et
al. 1999).  The EWs turned out to be larger than found here for M81
($360^{+175}_{-135}$ eV, and $572^{+378}_{-357}$ eV respectively).

Also the absence of rapid variability on timescales less than a day, a
likely case for the nucleus of M81 (Petre et al. 1993), seems to be a
general property of the LINERs observed with {\it ROSAT} and $ASCA$,
while changes in the X-ray luminosity over long timescales are not
uncommon (Reichert et al. 1994, Ptak et al. 1998). This has been
interpreted in terms of an ADAF being at work (Ptak et al. 1998).

\section { Conclusions}				

The (0.1--100) keV study of M81 based on $BeppoSAX$ data has revealed
the presence of a power law component, of emission and absorption
features due to highly ionized iron, and short term variability with a
period of $\sim 2 $ days.  These characteristics point to the presence
of a LLAGN, as suggested by I96 based on $ASCA$ data.  Differences and
similarities with the X-ray spectra of more luminous Seyfert galaxies
have been investigated.

The main result of the present work is that a 6.7 keV emission line,
corresponding to K$\alpha $ emission from He-like iron, is clearly
detected (confirming a previous $ASCA$ result) together with an
absorption edge at $\sim 8.6$ keV, due again to highly ionized iron.
The ionization level is consistent for the line and the edge, and
perhaps the same is true also for the column density of ionized
material.  So, we suggest that line and edge are produced by a highly
ionized absorber, that could be located within $ 10^{16}$ cm from the
source of ionizing photons.  Such an optically thin gas that transmits
the X-ray photons represents a similarity with the warm absorber in
Seyfert 1's.

Concerning the main spectral property of the continuum, $BeppoSAX$
data reveal the presence of a single power law of photon index
$\Gamma\sim 1.84$ in the (0.1--100) keV energy band.  This is modified
by cold absorption of $N_{\rm H}\sim 10^{21}$ cm$^{-2}$, that could be
entirely produced by the Galactic hydrogen column density plus some
intrinsic $N_{\rm H}$ corresponding to the reddening observed for the
broad line region.  The $\Gamma$ value, and the absence of high intrinsic
absorption, suggest that M81 might be an example of a low luminosity
Seyfert 1.  However, there are also differences with the X-ray
properties of Seyfert 1's: there is no sign of reflection from
optically thick cold material, like the 6.4 keV emission line and the
broad bump peaking at 10--20 keV. Even a blue bump is absent in the
spectral energy distribution of the nucleus of M81, all this
suggesting that this nucleus cannot be considered a simple extension to low
 luminosities of classical Seyfert 1's.

Alternatively to the standard accretion disk scenario, the continuum
could be produced by an ADAF. This hypothesis is consistent with the
low accretion rate inferred from the luminosity and the estimate of
the central black hole mass; it remains to be investigated whether
the ADAF scenario can reproduce a Seyfert-like power law with $\Gamma
\sim 1.84$ over 0.1--100 keV for M81.

\begin{acknowledgements}
This research has made use of SAXDAS linearized and cleaned event
files (rev0) produced at the BeppoSAX Science Data Center. We are
grateful to G. Fabbiano and N. Iyomoto for showing us their
results in advance of publication, and to  H. Netzer, F. Nicastro
and T. Di Matteo for discussions. 
ASI and MURST (contract CoFin98) are acknowledged for financial support.
\end{acknowledgements}

\end{document}